*Original Paper*

# Reciprocal Phase Transition Electro-optic Modulation

*Fang Zou,[1,†] Lei Zou,[2,†] Ye Tian,[3] Yiming Zhang,[4] Erwin Bente,[3] Weigang Hou,[5] Yu Liu,[4] Siming Chen,[2] Victoria Cao,[2] Lei Guo,[5] Songsui Li,[1] Lianshan Yan,[1] Wei Pan,[1] Dusan Milosevic,[3] Zizheng Cao,[3,6,*] A. M. J. Koonen,[3] Huiyun Liu,[2] and Xihua Zou[1,*]*

[1] School of Information Science and Technology, Southwest Jiaotong University, Chengdu 611756, China.
[2] Department of Electronic and Electrical Engineering, University College London, WC1E 6BT London, UK.
[3] Institute for Photonic Integration, Eindhoven University of Technology, Eindhoven 5600 MB, Netherlands.
[4] Institute of Semiconductors, Chinese Academy of Sciences, Beijing 100083, China.
[5] School of Communication and Information Engineering, Chongqing University of Posts and Telecommunications, Chongqing 400000, China.
[6] Peng Cheng Laboratory, Shenzhen 518055, China.

† The authors contributed equally to this work.
*Corresponding Author: E-mail: zouxihua@swjtu.edu.cn (X.Z.), z.cao@tue.nl (Z.C.).

**Abstract**: Electro-optic (EO) modulation is a well-known and essential topic in the field of communications and sensing, while ultrahigh modulation efficiency is unprecedentedly desired in the current green and data era. However, dramatically increasing the modulation efficiency is difficult in conventional mechanisms, as being intrinsically limited by the monotonic mapping relationship between the electrical driving signal and modulated optical signal. To break this bottleneck, a new mechanism termed phase-transition EO modulation is revealed from the reciprocal transition between two distinct phase planes arising from the Hopf bifurcation, being driven by a transient electrical signal to cross the critical point. A monolithically integrated mode-locked laser is implemented as a prototype, strikingly achieving an ultrahigh modulation energy efficiency of 3.06 fJ/bit improved by about four orders of magnitude and a contrast ratio exceeding 50 dB. The prototype is experimentally implemented for radio-over-fiber communication and acoustic sensing. This work indicates a significant advance on the state-of-the-art EO modulation technology, and opens a new avenue for green communication and ubiquitous sensing applications.

## 1. Introduction

Electro-optic (EO) modulators are at the heart of optical communication, microwave photonic, and sensing systems for implementing EO conversion. Over the past decades, large amounts of modulators have been developed which underpins the advances of modern modulation technology, including phase, intensity, electronic absorption, and polarization modulators [1-8]. In the current green and data era, the demand of high-efficiency, high-stability, and compact EO modulation remains of paramount importance for communication and sensing.

A common intrinsic feature among conventional EO modulations (including direct and external cases) is the monotonic linear or nonlinear mapping between the applied electrical driving signal and the modulated light, wherein the modulation states are located in the same phase plane with identical physical dynamics or equations. Thus, their modulation efficiency is restricted by the EO response of materials and structures [9-12]. Specifically, the modulation efficiency is defined as the characteristic change in the light signal versus the value (current, voltage, or power) of the applied electrical driving signal. In addition, the pattern of the output light signal is the analogue of the electrical driving signal due to the monotonic mapping relationship in conventional modulations. For instance, EO modulation of the radio-over-fiber system where the intermediate frequency (IF) or radio-frequency (RF) signal serves as the driving signal tends to increase power consumption, and the modulation efficiency is generally estimated as pJ/bit or sub-pJ/bit [11, 12]. In this paper, a new mechanism termed phase transition EO modulation is proposed by utilizing the reciprocal cross transition between distinct phase planes in a dynamical system. Remarkably, a monolithically integrated multi-section mode-locked laser is developed as a prototype to demonstrate the phase transition EO modulation, which is used in radio-over-fiber communication and acoustic sensing systems.

## 2. Concept of phase-transition EO modulation

To describe the EO modulation mechanism, the DML is taken as an object having a monotonic linear modulation (or mapping) response beyond the threshold current, as shown in **Figure 1a**. The emitted optical power rigorously follows the amplitude change of the applied electrical driving signal in accordance with the monotonic linear mapping response, realizing EO modulation. The modulation efficiency of the DML is determined by the laser slope efficiency with an identical physical equation. Other modulation mechanisms, such as those in PMs, MZMs or electro-absorption modulators (EAMs), also entail a similar process in which the electrical driving signal modifies the modulated optical signal in the same phase plane. In these conventional EO modulation mechanisms, the modulation efficiency is mainly limited by the monotonic mapping response in principle and is determined by the materials and structures of the modulators. Moreover, the monotonic mapping response occurs in the same phase plane. When a square wave or a sine wave serves as a driving electrical signal, a modulated optical signal with the same pattern is generated in the optical domain, as depicted in **Figure 1c**. A LO carrying the baseband signal is necessary for microwave or wireless transmission links (e.g., Radio-over-fiber system).

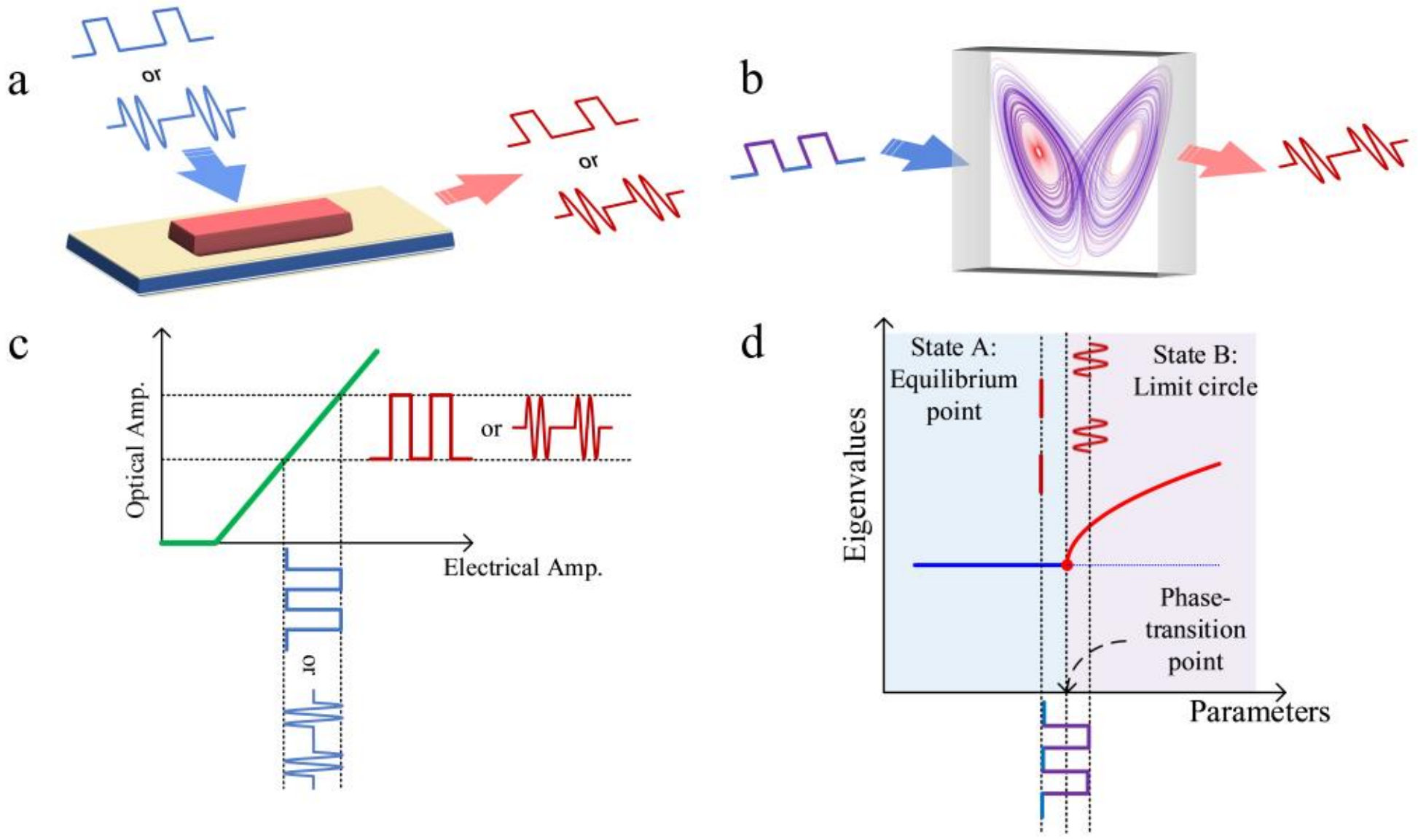

Figure 1. Comparison of EO modulation mechanisms. (a) The conventional EO modulation is considered an electrical-to-optical mapping following a continuous curve inside a single plane. (b) The proposed phase-transition EO modulation is regarded as an electrical-to-optical mapping among discrete and isolated planes. (c) A DML with a linear modulation response is considered to correspond to the conventional EO modulation. (d) The mapping relationship between the electrical driving and the output optical signals in phase-transition modulation. When a tiny electrical signal drives the parameters across the phase-transition point between two phase planes, the dynamical system will spontaneously evolve to the reciprocal state in the other phase plane with distinct characteristics.

If we describe the EO modulation using a dynamical system with bifurcation, then the modulation mechanism is highly divergent from the conventional EO modulation, as shown in **Figures 1b** and **d**. The dynamical system can be modelled using differential equations, the initial parameter values of which result in distinct behaviors (i.e., solutions). When the value of a parameter exceeds the phase-transition point, the trajectory structures of the phase planes are different, including the sole equilibrium, multiple equilibria, or limit cycle. The differential equations have bifurcations among these discrete and isolated phase planes. Hence, a tiny driving signal across the phase-transition point enables switching of the dynamical system between discrete phase planes.

Specifically, the two-dimensional dynamical system with Hopf bifurcation is taken as an example to elaborate the new modulation mechanism [13-15]:

$$\dot{r} = r(\mu - r^2) \quad \textbf{(1)}$$

where $r$ is a vector in polar coordinate and $\mu$ is a control parameter. As $\mu$ crosses the critical phase transition point $\mu_c = 0$, bifurcation occurs and a periodic solution or limit cycle surrounding the unstable equilibrium point arises or vanishes, known as the supercritical Hopf bifurcation shown in **Figure 2a**.

When a driving signal sets the control parameter as $\mu < \mu_c$, the eigen solution of the dynamical system exhibits a stable equilibrium. The trajectory structure of the phase plane spontaneously evolves to a stable focus (i.e., State A) from the initial state, as shown in **Figure 2b**. If the parameter is configured as $\mu > \mu_c$, a periodic eigen solution is derived with a stable orbit, as shown in the phase portrait in **Figure 2c**. The solution will diverge from the original focus until approaching a stable orbit if the initial state is smaller than the limit cycle. Otherwise, the solution will converge to a stable orbit from the outside of the limit cycle. Anyway, the initial state will spontaneously evolve to a stable orbit (i.e., State B). Thus, the dynamical system spontaneously evolves to the self-sustaining State A or B.

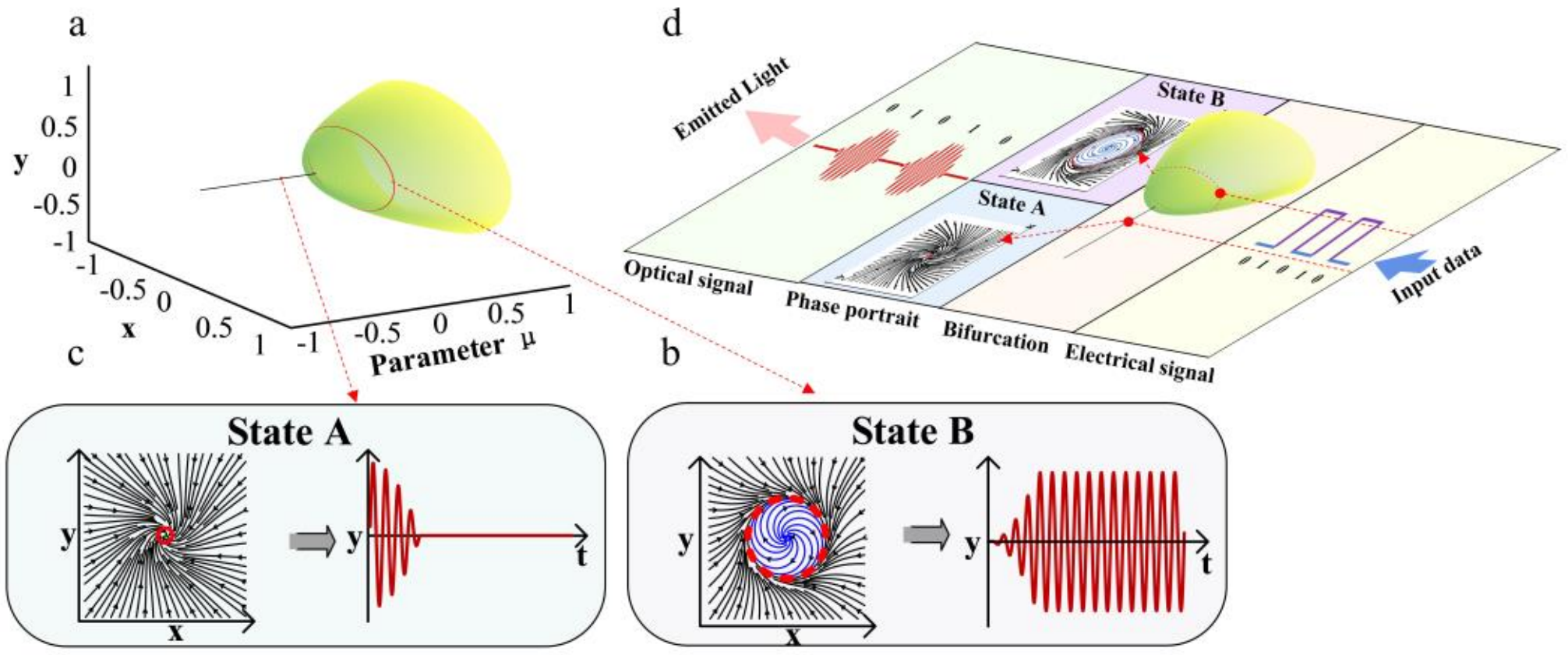

Figure 2. Principle of the phase transition EO modulation. (a) Dynamical system with supercritical Hopf bifurcation. Phase portrait and vector of (b) State A and (c) State B: the trajectory structure of the phase portrait spontaneously evolves to equilibrium (State A) for $\mu < \mu_c$ or to a stable orbit (State B) for $\mu > \mu_c$. (d) Procedure overview of the phase transition EO modulation.

According to the supercritical Hopf bifurcation of the dynamical system, the phase transition EO modulation can be realized by apply a tiny driving signal to cross the critical point $\mu = \mu_c$, as shown in **Figure 2d**. Since the characteristics of the two modulation states (States A and B) are totally distinct, a high contrast ratio between them is achieved by applying a tiny driving signal. In addition, the output of the dynamical system has an intrinsic RF frequency carrying the baseband driving signal. Therefore, the generation and modulation of RF signal is simultaneously achieved through this phase transition EO modulation, with no need of local oscillator and mixer for up-conversion. Meanwhile, the phase transition EO modulation is strongly immune to fluctuations in the driving signal. Because the modulation states can remain self-sustaining as long as the fluctuations on the driving signal do not exceed the phase transition point.

Further, semiconductor lasers with rich nonlinear dynamics [16] such as injection, feedback and mode-locking, can be adopted to implement the proposed phase transition EO modulation. Here a monolithically integrated multi-section mode-locked laser is designed and fabricated to reveal the phase transition features, using InP photonic integration technology [17].

## 3. Mechanism of phase-transition EO modulation in the MLL

A semiconductor multi-section mode-locked laser is designed to reveal the new modulation mechanism, as shown in **Figure 3a.** Sections I and III are a gain element (i.e., semiconductor optical amplifier, SOA) and a passive section, respectively. Section II connected to a switching port is a reconfigurable element (i.e., a short SOA). At the interface between Sections I and III, a part of the light is coupled out through the transmission window of the distributed Bragg reflector (DBR), while the remaining light is reflected by the DBR into the ring cavity.

When a current exceeding the threshold is injected into the gain medium, a light will emit and the bifurcation of the mode-locked laser is determined by the value of the control parameter, $\mu$. Here, a new method to adjust the control parameter's value is discovered, namely the alternative transparency and saturable absorption effects in a reconfigurable element (i.e., Section II). Specifically, the laser state and the emitted light state are manipulated by Section II connected to a switching port. When the switching port is triggered to be open-circuit, Section II behaves as an absorption element with linear loss. After a temporal period for the phase transition to take place, the mode-locked laser evolves to a self-sustaining continuous-wave (CW) state (i.e., State A) and emits a CW light, as shown in **Figure 3b**. In contrast, when the switching port is set to be grounded, part of the oscillating light energy transforms into an electrical voltage being applied on Section II. Owing to the stimulated absorption, Section II serves as a saturable absorber (SA) in the mode-locked laser cavity [18]. The mode-locked laser spontaneously evolves to the self-sustaining pulsed-wave (PW) state (i.e., State B) and emits pulsed light trains, as shown in **Figure 3c**. Hence the mode-locked laser spontaneously evolves between the CW or PW states and thus emits a CW light or pulsed trains, when the electrical signal triggers transient actions through a control circuit to set the switching port as open-circuit or grounded, as shown in **Figure 3d**. Note that a few switches triggered by a tiny and transient electrical signal can serve as the control circuit, such as the floating-gate MOSFET switch. Thanks to the use of a transient pulse, the modulation power is greatly reduced in this phase transition EO modulation.

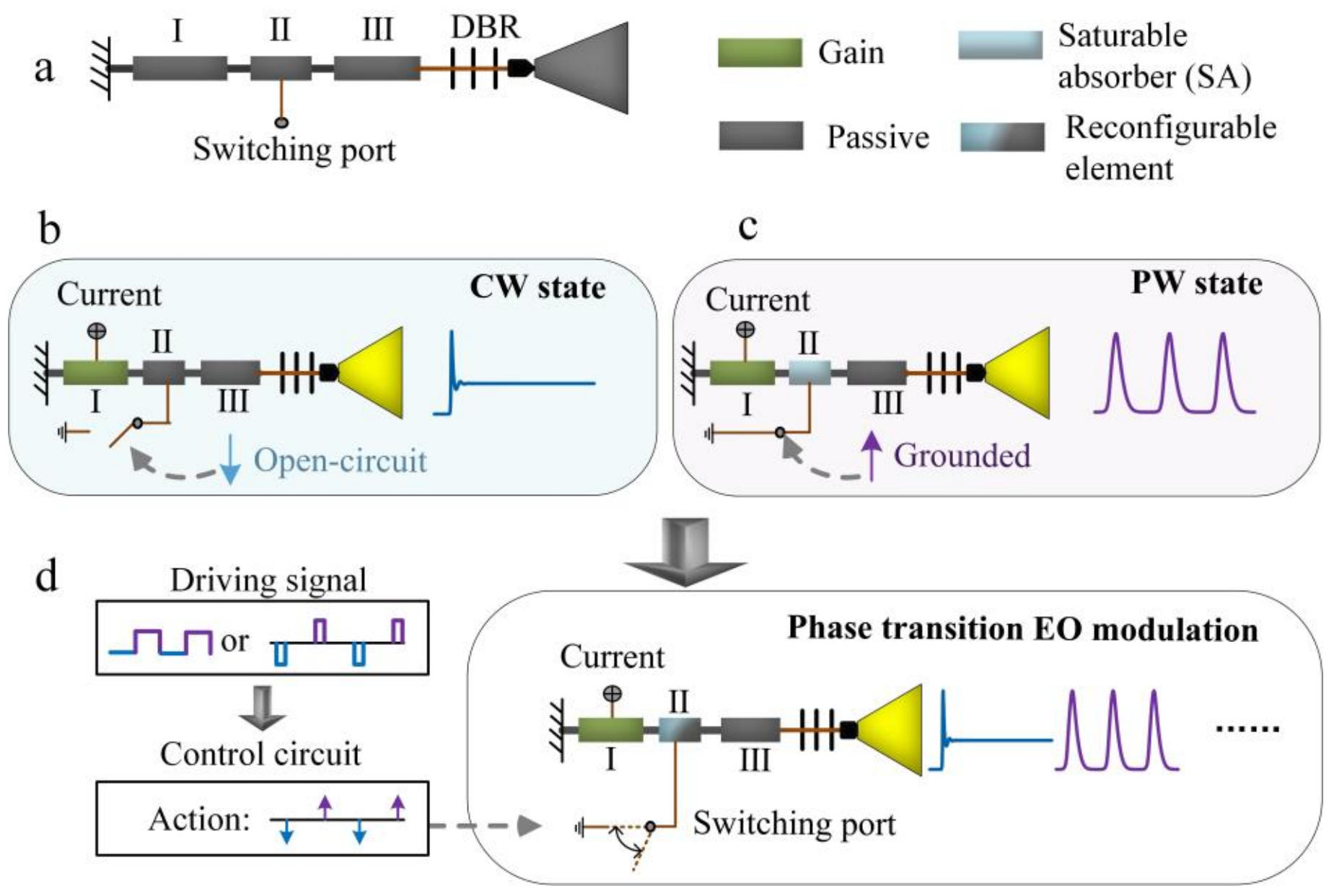

Figure 3. Phase transition EO modulation in mode-locked laser model. (a) Multi-section mode-locked laser. (b) CW state and (c) PW state achieved by configuring the SA connected to a switching port. (d) Driving signal and control circuit for switching between open-circuit and grounded actions.

The delay differential equation (DDE) model with a ring cavity is usually used to investigate the qualitative dynamics for mode-locked lasers with linear cavity [19-29]. The differences between the travelling-wave model for a linear cavity and the DDE model for a ring cavity are discussed in Section S1, Support information. For the CW state (i.e., “0”), the stimulated absorption is absent in the PN junction of the Section II. A single-tone CW light is emitted, whose electrical spectrum shows almost noise after the DC blocking, as shown in **Figures 4a** and **b**. When the switching port is grounded, Section II serves as an SA that periodically absorbs photons and becomes transparent owing to the stimulated absorption. The mode-locked laser spontaneously evolves to the self-sustaining PW state such that pulsed trains are generated. The corresponding optical and electrical spectra are shown in **Figures 4c** and **d**, indicating frequency combs with a spacing at the pulse repetition rate. The parameters, processes, results, and evolution of the DDE model are detailed in Section S2 and S3, Support information.

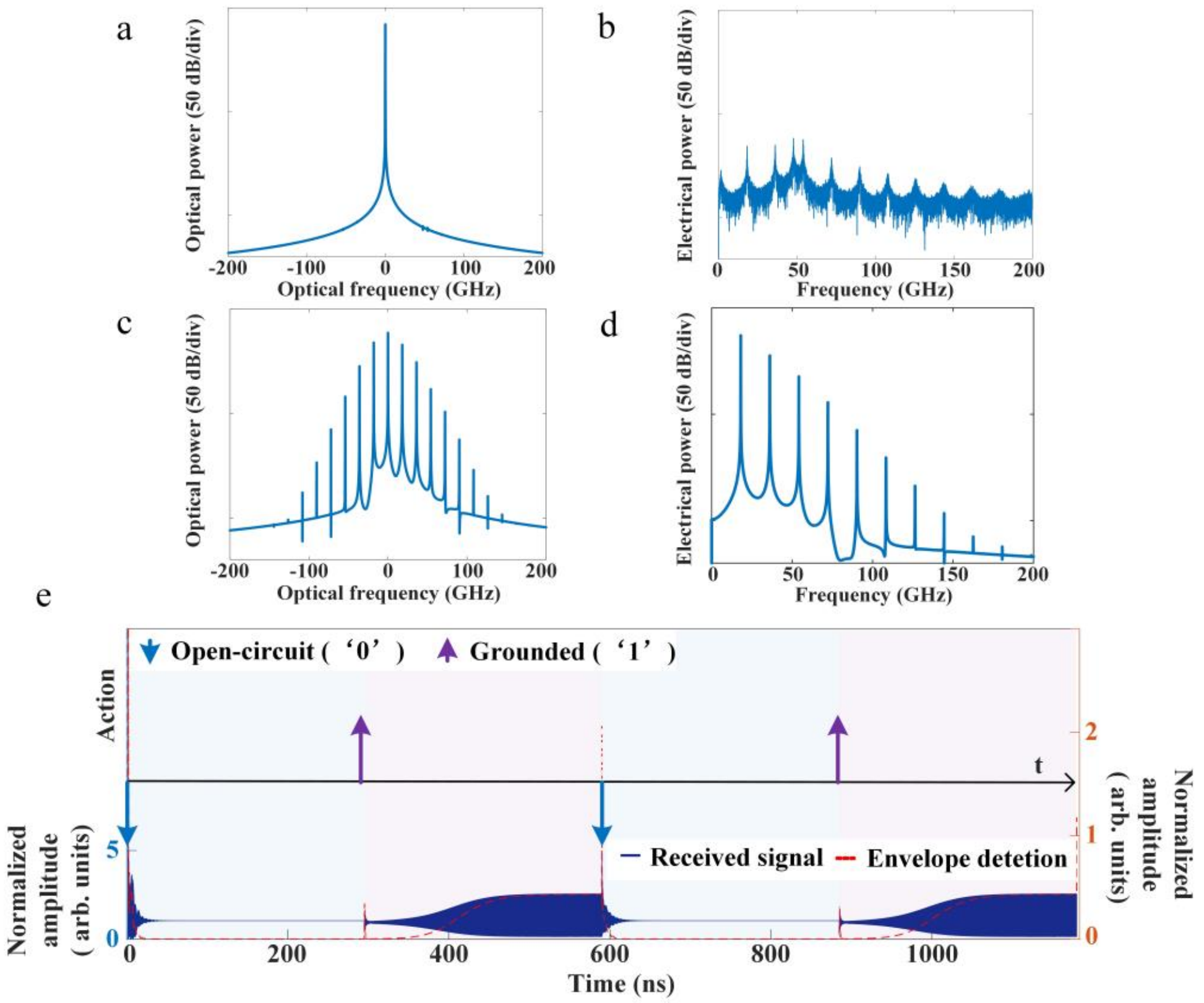

Figure 4. Simulation results of the phase transition EO modulation. (a) Optical and (b) electrical spectra of the PW state. (c) Optical and (d) electrical spectra of the CW state. (e) Qualitative dynamics of spontaneous CW-PW and PW-CW state transitions. In the blue range, the baseband signal is coded as "0" at the CW state, while in the purple range the signal is coded as "1" at the PW state.

During the EO modulation, a coded baseband signal serves as the driving to trigger the switching port via the control circuit. In the simulation, the coded signal "1" or "0" is converted into the grounded or open-circuit action with a period of 294.912 ns. Under the grounded action, the mode-locked laser works at the PW state and emits pulsed light trains. In contrast, the mode-locked laser being configured at the open-circuit port is operating at the CW state. The simulation results of the phase transition EO modulation based on the mode-locked laser is illustrated in **Figure 4e**. The transition times are approximately 200 ns and 40 ns for the spontaneous CW-PW and PW-CW state transitions, respectively. At the remote receiver, the intrinsic frequency carrying the coded baseband signal is directly generated, and then the digital code can be recovered via envelope detection, indicated by the orange dashed line in **Figure 4e**. Moreover, the intrinsic frequency of the mode-locked laser has the potential to cover an extensive frequency range from MHz to tens GHz [30]. In this way, RF up-conversion with a local oscillator is unnecessary when using this phase transition EO modulation, resulting in a simplified radio-over-fiber system and ultralow modulation power consumption.

## 4. Realization and characterization of the MLL protype

To realize the phase transition EO modulation, a monolithic mode-locked laser is fabricated as a protype, via InP technology offered by the JePPIX platform [17]. As shown in **Figure 5**, it has a multi-section linear cavity consisting of a multimode interference reflector (MIR), a rear DBR (DBR-R), a long SOA (SOA-G), a short SOA (SOA-S), a phase shifter (PS) and a front DBR (DBR-F). The MIR and the 200-μm DBR-F form a linear cavity with a total length of 1580 μm. The length of the long SOA (i.e., SOA-G) is 500 μm, which serves as a gain element to provide sufficient gain for laser oscillation. Meanwhile, the short SOA (i.e., SOA-S) with a length of 50 μm is connected with a switching port, which can be reconfigured as SA and a passive element with linear loss. Here, the SOA-G corresponds to Section I and SOA-S being connected to a switching port refers to Section II in the DDE model respectively, while other passive waveguides inside the laser cavity can be equivalently grouped as Section III.

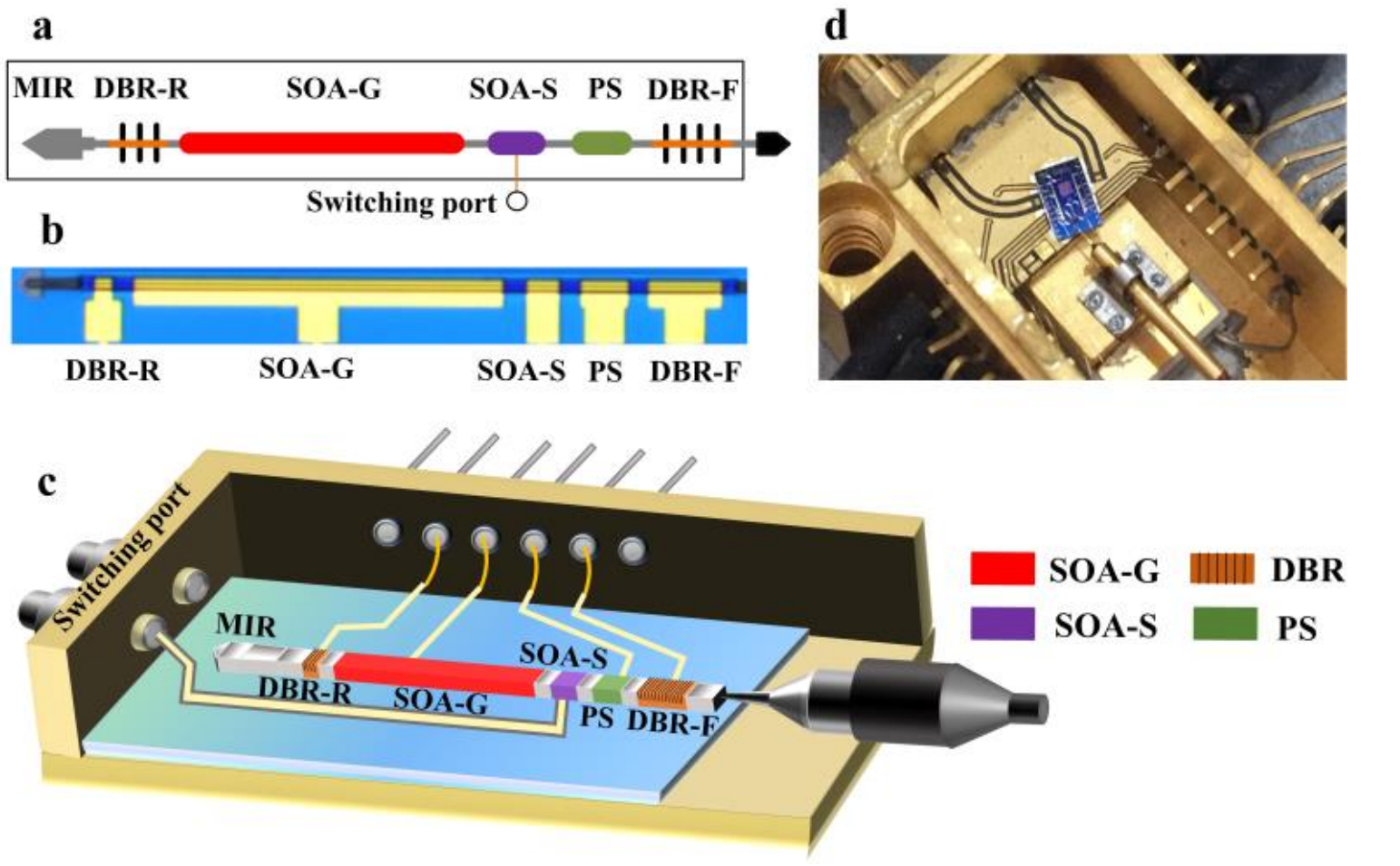

Figure 5. Fabricated mode-locked laser protype for implementing the phase transition EO modulation. (a) Layout and (b) microscopic image of the mode-locked laser. (c) Schematic diagram and (d) photo of the packaged mode-locked laser.

When a current is injected into the SOA-G, a light is emitted from the monolithic mode-locked laser and its state can be tuned by switching the SOA-S to be grounded or open-circuit. The blue line in **Figure 6a** shows the optical spectrum of the self-sustaining CW state when the SOA-S is configured as a linear-loss element (i.e., open-circuit). The generated CW light is centered at 1551.2 nm. In contrast, the red line in **Figure 6a** shows the optical spectrum of the self-sustaining PW state when the SOA-S is reconfigured to be grounded. The generated pulsed light trains consists of multiple correlated comb lines with a frequency spacing of 24.8 GHz. To distinguish the characteristics of the CW and PW states, the emitted light is converted into electrical signal by a photodetector and then analyzed. As indicated by the blue line in **Figure 6b**, only noise appears in the electrical spectrum of the CW state after DC blocking. On the other hand, the pulsed light trains generated from the PW state are converted into an RF signal after photodetection. Such an RF signal with a fundamental frequency of 24.8 GHz is depicted as the red line of **Figure 6b**, showing a contrast ratio over 50 dB between the CW and PW modulation states. Furthermore, tunable wavelength and fundamental RF frequency are enabled in this laser protype. As depicted in **Figure 6c**, the fundamental frequency of the generated RF signal is tuned from 24.8 to 23.7 GHz for versatile applications. The phase noise of the generated RF signal is measured to be lower than -100 dBc/Hz @ 1-MHz offset frequency, as shown in **Figure 6d**.

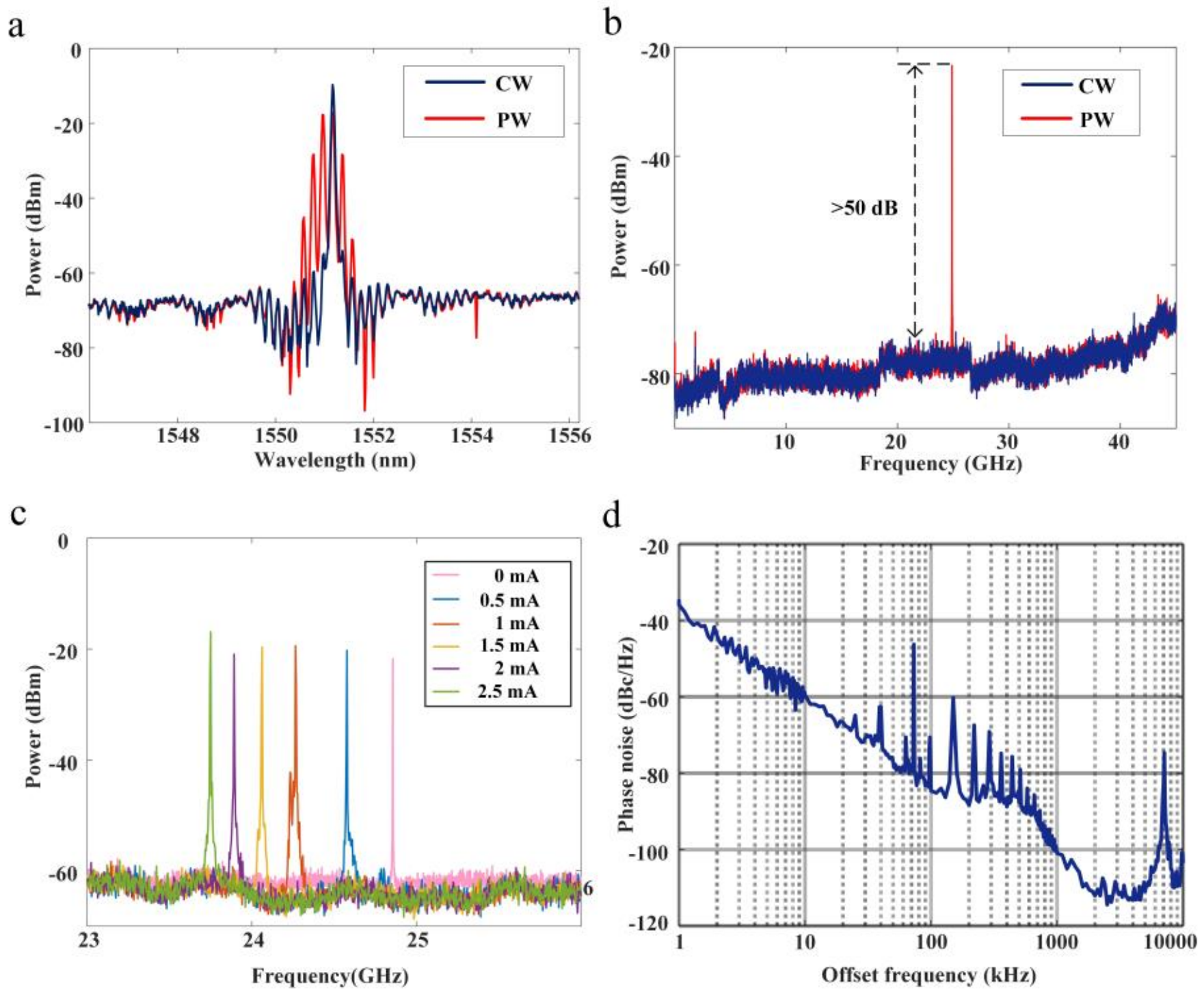

Figure 6. Characteristics of the CW and PW states in the mode-locked laser protype. (a) Optical and (b) electrical spectra. (c) Tunable fundamental frequency and (d) phase noise of the generated RF signal.

Due to the ultrahigh modulation efficiency, a high contrast ratio between states can be realized by using a tiny driving electrical signal. Here a MOSFET switch driven by a 70-mV electrical signal serves as the control circuit to trigger the grounded or open-circuit action. The modulation energy efficiency is calculated to be 3.06 fJ/bit based on the driving voltage and the capacitance of the MOSFET. Compared with the modulation energy efficiency of pJ/bit or sub pJ/bit in conventional EO modulations [31], the proposed phase transition EO has the value enhanced by four orders of magnitude. Such an efficiency can be enhanced further by using a less capacitance. The calculation of the modulation energy efficiency is discussed in in Section S4, Support information.

## 5. Application Demonstrations

### *5.1. Radio-over-fiber communication*

Using the mode-locked laser protype as a compact transmitter, a radio-over-fiber communication system is designed and described in **Figure 7a**. A 2-Mbit/s baseband signal triggers the switching port to implement EO modulation. The modulated light is transmitted over a 10-km single-mode fiber to a remote unit. After being detected by a photodetector, the pulsed light trains are converted as a modulated RF signal at 24.8 GHz. On the other hand, the CW light is detected as a DC component and then blocked. Thus, a 24.8-GHz RF signal carrying the 2-Mbit/s amplitude-shift-keying (ASK) baseband signal is generated in the remote radio unit and radiated out by antenna for wireless communication. After a 1.8-m transmission in free space, the RF signal is received by another antenna, and then the 2-Mbit/s ASK signal is recovered by using an electrical envelope detector. **Figures 7b** and **c** displays the temporal waveform and the bit error rates (BER) curve of the recovered baseband signal for the radio-over-fiber system. Assuming a hard-decision forward error correction (FEC) with a threshold of BER≤$3.8\times10^{-3}$, the error-free received optical power needs to exceed -9.2 dBm at the remote unit.

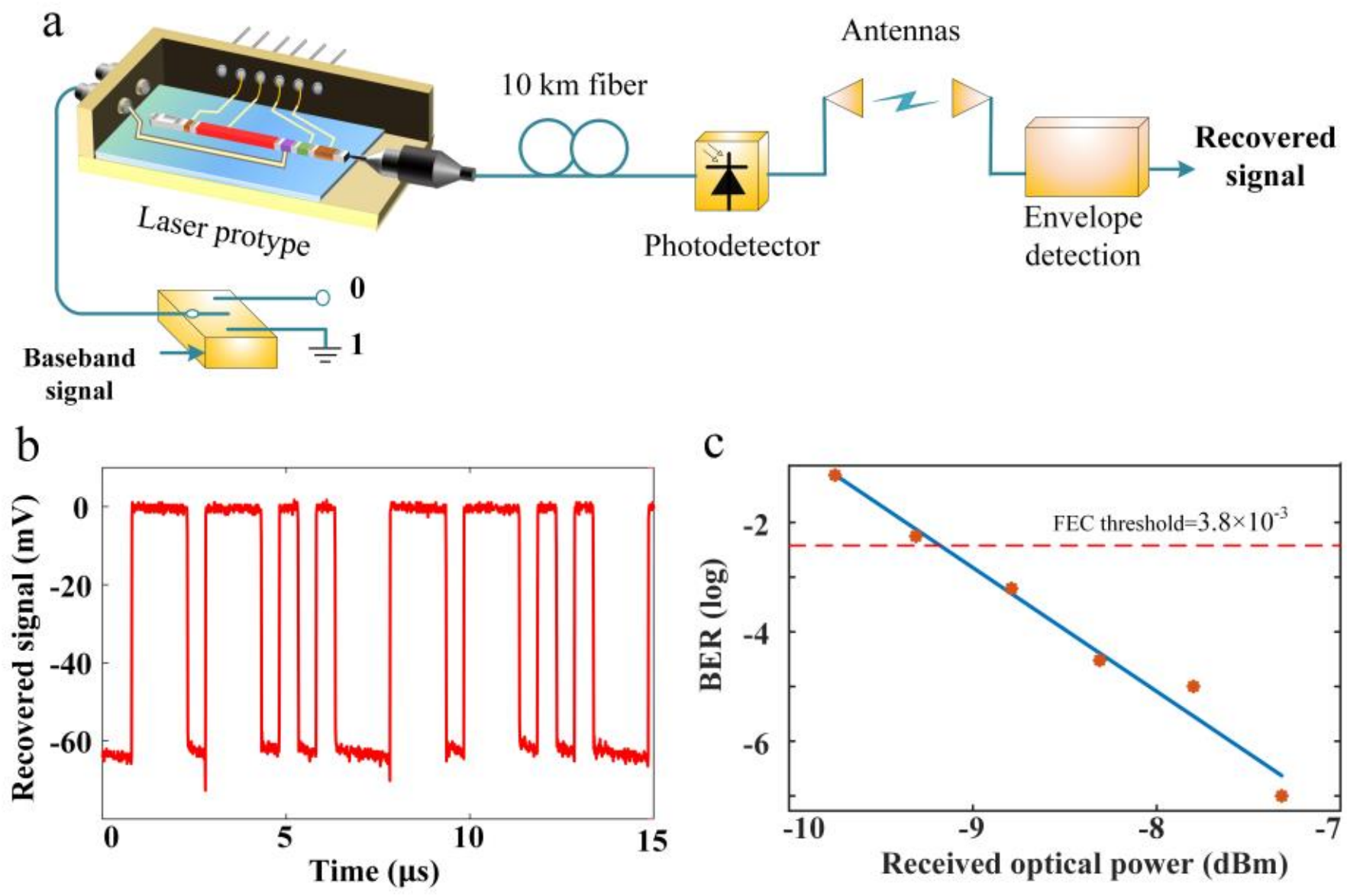

Figure 7. Radio-over-fiber communication. (a) System diagram. (b) Temporal waveform and (c) BERs of the recovered baseband signal.

### *5.2. Underwater acoustic sensing.*

Underwater acoustic sensing networks were intensively investigated by integrating acoustic and RF techniques [32]. However, their combination is insufficient to overcome obstacles in specific circumstances, e.g., a large underwater gap between the deep-water acquisition points of weak acoustic signals and the atmosphere surface for RF transmission. Fortunately, the mode-locked laser protype allows the gap to be filled with a simple scheme for this deep-water sensing system, with economic power consumption through the full integration of acoustic, RF, and optical techniques.

As shown in **Figures 8a and b**, the laser protype serves as a receiving relay and transmits the underwater acoustic signal captured by a local hydrophone. A 200-kHz acoustic signal carrying a 100 kbit/s baseband signal is generated. The acoustic signal transmits to the relay module inside a water pool and is detected by a local hydrophone. Then, the received acoustic signal is processed and converted into the grounded or open-circuit actions via a MOSFET. After the phase transition EO modulation implemented by the mode-locked laser protype, the emitted light signals are transmitted through a 5-km fiber to the water surface. Then a 24.8-GHz RF signal carrying a 100-kbit/s baseband signal is detected and radiated out to establish a wireless communication link. Subsequently, the RF signal is received by an antenna on the coast, and the baseband signal is recovered through electrical envelope detection. The recovered signal and the corresponding BERs are depicted in **Figures 8c-e**. Consequently, the minimum achievable received optical power is around -5 dBm for this underwater acoustic sensing application, assuming a hard-decision FEC with a BER threshold no more than $3.8\times10^{-3}$.

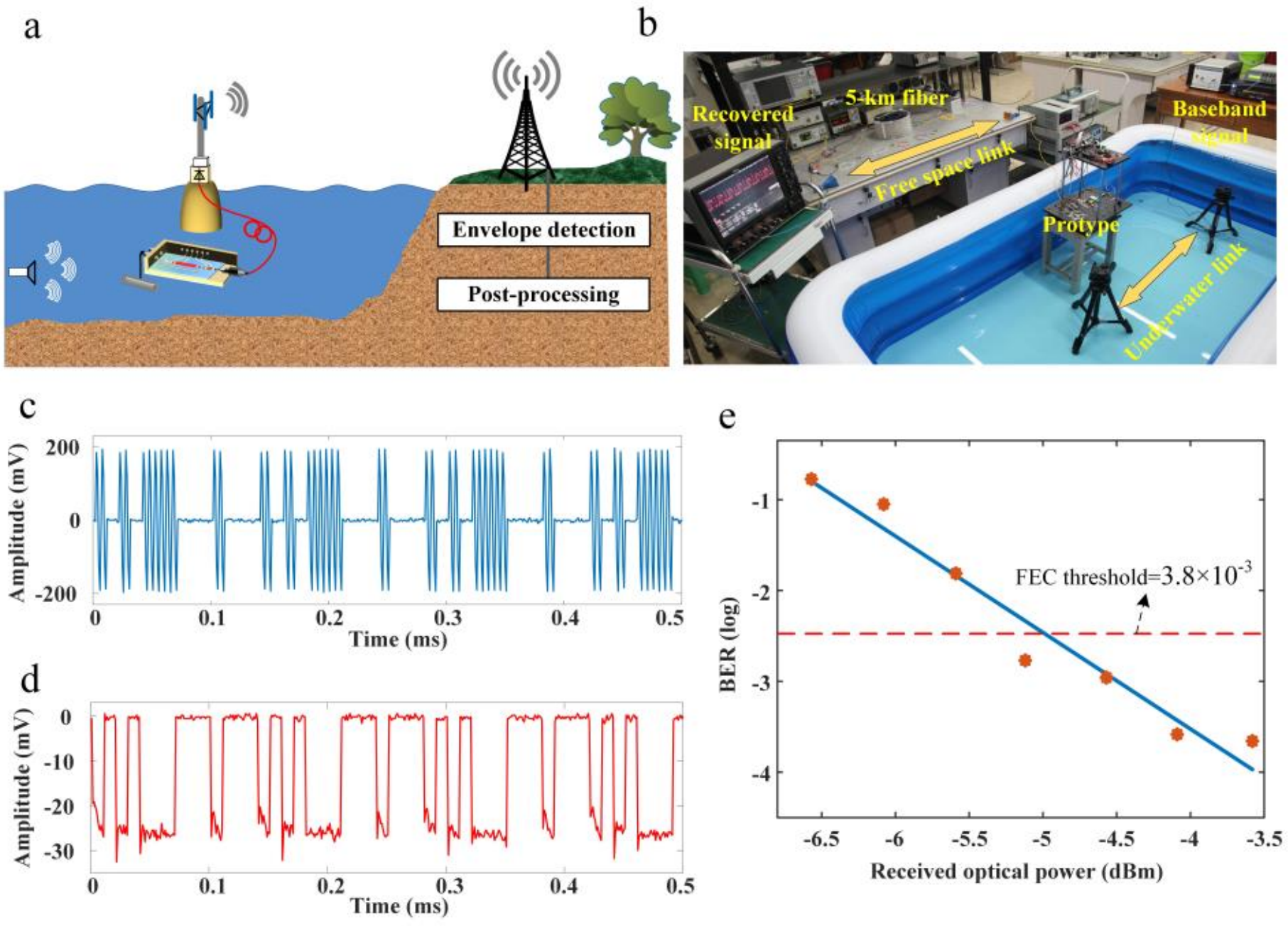

Figure 8. Underwater acoustic sensing. (a) System diagram and (b) experimental setup. (c) Transmitting signal: 100 kbit/s baseband signal carried by acoustic signal. (d) Temporal waveform and (e) BERs of the recovered baseband signal.

## 6. Conclusion and Discussions

The new phase transition EO modulation is proposed to break the modulation efficiency bottleneck in the conventional EO modulators. During the proposed EO modulation, a tiny driving signal around the critical phase transition point is instantaneously applied to trigger the dynamical bifurcation system, leading to spontaneous evolution to self-sustaining CW or PW state. Such a phase transition EO modulation is demonstrated through an integrated mode-locked laser protype. The modulation energy efficiency is improved by four orders of magnitude to be 3.06 fJ/bit, being compared with that of the conventional EO modulations. Meanwhile, the ultrahigh contrast ratio exceeds 50 dB for the simultaneous generation and modulation of RF signals. As two application examples, the integrated protype of the phase transition EO modulation has been successfully demonstrated in radio-over-fiber communication and acoustic sensing scenarios.

Furthermore, the nonlinear dynamics optimization to the phase transition in mode-locked lasers and the use of high-order modulation formats (like PAM4 or PAM8) are able to realize higher data rate for communication and sensing applications [33, 34]. The reconfigurable multi-section MLL can be considered to realize the high-order modulation. Different dynamics of high-order harmonics would serve as a series of self-sustaining states via reconfiguring the multi-section cells, including the gain elements, saturable absorbers and passive sections [33]. In the remote receiver, a M-ary frequency-shift keying (FSK) signal is generated, providing the capability of high-order modulation.

Thus, this new phase transition EO modulation mechanism and its protype are featured by high modulation efficiency, high stability, and high contrast ratio for green communications and ubiquitous connections in the foreseeable future.

### Supporting Information

Supporting Information is available from the Wiley Online Library or from the authors.

**Section S1.** Qualitative differences between the travelling-wave model for a linear cavity and the DDE.

**Section S2.** Theoretical analysis of phase transition EO modulation in the mode-locked laser based on the DDE model.

**Section S3.** Parameters, processes, and results of the simulation.

**Section S4.** Calculation of modulation energy efficiency.

**Acknowledgments**

F.Z., and L.Z. contributed equally to this work. This work is partly supported by the National Natural Science Foundation of China (61922069, U21A20507), Sichuan Provincial Natural Science Foundation (23NSFC0425), NWO Zwaartekracht program on Integrated Nanophotonics, the Major Key Project of PCL (PCL2021A14), and the Engineering and Physical Sciences Research Council (EP/T01394X/1).

**Conflict of Interest**

The authors declare no conflict of interest.

**References**

[1]. Y. Matsui, R. Schatz, D. Che, F. Khan, M. Kwakernaak, and T. Sudo, *Nat. Photon.* **2021**, *15*, 59-63.

[2]. N. N. Feng, D. Feng, S. Liao, X. Wang, P. Dong, H. Liang, C. C. Kung, W. Qian, J. Fong, and R. Shafiiha, *Opt. Express* **2011**, *19*, 7062-7067.

[3]. G.T. Reed, G. Mashanovich, F. Y. Gardes, and D. J. Thomson, *Nat. Photon.* **2010**, 4, 518-526.

[4]. M. He, M. Xu, Y. Ren, J. Jian, Z. Ruan, Y. Xu, S. Gao, S. Sun, X. Wen, and L. Zhou, *Nat. Photon.* **2019**, *13*, 359-364.

[5]. P. Kolchin, C. Belthangady, S. Du, G. Y. Yin, and S. E. Harris, *Phys. Rev. Lett.* **2008**, *101*, 103601.

[6]. W. Yao, M. K. Smit, and M. J. Wale, *IEEE J. Sel. Top. Quantum Electron.* **2017**, *24*, 6100711.

[7]. M. Li, L. Wang, X. Li, X. Xiao, and S. Yu, Photonics Res. 2018, **6**, 109-116.

[8]. S. Liu, P. C. Peng, M. Xu, D. Guidotti, H. Tian, and G.-K. Chang, *IEEE Photon. Technol. Lett.* **2018**, *30*, 1396-1399.

[9]. H. Yu, M. Pantouvaki, J. Van Campenhout, D. Korn, K. Komorowska, P. Dumon, Y. Li, P. Verheyen, P. Absil, and L. Alloatti, *Opt. Express* **2012**, *20*, 12926-12938.

[10]. M. T. Rakher, L. Ma, M. Davanço, O. Slattery, X. Tang, and K. Srinivasan, *Phys. Rev. Lett.* **2011**, *107*, 083602.

[11]. P. Dong, S. Liao, D. Feng, H. Liang, D. Zheng, R. Shafiiha, C.-C. Kung, W. Qian, G. Li, and X. Zheng, *Opt. Express* **2009**, *17*, 22484-22490.

[12]. S. S. Azadeh, F. Merget, S. Romero-García, A. Moscoso-Mártir, N. von den Driesch, J. Müller, S. Mantl, D. Buca, and J. Witzens, *Opt. Express* **2015**, *23*, 23526-50.

[13]. S. H. Strogatz, *Nonlinear dynamics and chaos: with applications to physics, biology, chemistry, and engineering*, (CRC press, **2018**).

[14]. J. E. Marsden and M. McCracken, *The Hopf bifurcation and its applications*, (Springer Science & Business Media, **2012**).

[15]. Y. A. Kuznetsov, *Elements of applied bifurcation theory*, (Springer Science & Business Media, **2013**).

[16]. T. J. Kippenberg, R. Holzwarth, and S.A. Diddams, *Science*, **2011**, *332*, 555-559.

[17]. X. Leijtens, *IET Optoelectron*. **2011**, *5*, 202-206.

[18]. S. M. Sze and K. K. Ng, *Physics of semiconductor devices*, (John Wiley & Sons, **2007**).

[19]. A. G. Vladimirov and D. Turaev, *Phys. Rev. A* **2005**, *72*, 033808.

[20]. L. Nielsen and M. J. A. Heck, *J. Lightwave Technol.* **2020**, *38*, 5430-5439.

[21]. L. Jaurigue, *Passively mode-locked semiconductor lasers: dynamics and stochastic properties in the presence of optical feedback*, (Springer, **2017**).

[22]. N. Rebrova, G. Huyet, D. Rachinskii, and A. G. Vladimirov, *Phys. Rev. E* **2011**, *83*, 066202.

[23]. A. G. Vladimirov, A. S. Pimenov, and D. Rachinskii, *IEEE J. Quantum Electron.* **2009**, *45*, 462-468.

[24]. J. R. Karin, R. J. Helkey, Dennis J. Derickson, R. Nagarajan, D. S. Allin, J. E. Bowers, and R. L. Thornton, *Appl. Phys. Lett.* **1994**, *64*, 676-678.

[25]. D. Zhao, S. Andreou, W. Yao, D. Lenstra, K. Williams, and X. Leijtens, *IEEE Photon. J.* **2018**, *10*, 6602108.

[26]. D. Pustakhod, K. Williams, and X. Leijtens, *IEEE J. Sel. Top. Quantum Electron.* **2017**, *24*, 3100309.

[27]. G. Liu, X. Jin, and S. L. Chuang, *IEEE Photon. Technol. Lett.* **2001**, *13*, 430-432.

[28]. G. Fiol, D. Arsenijević, D. Bimberg, A. G. Vladimirov, M. Wolfrum, E. A. Viktorov, and P. Mandel, *Appl. Phys. Lett.* **2010**, *96*, 011104.

[29]. U. Bandelow, M. Radziunas, A. Vladimirov, B. Hüttl, and R. Kaiser, *Opt. Quantum Electron.* **2006**, *38*, 495-512.

[30]. A. J. Ward, D. J. Robbins, G. Busico, E. Barton, L. Ponnampalam, J. P. Duck, N. D. Whitbread, P. J. Williams, D. C. Reid, and A. C. Carter, *IEEE J. Sel. Top. Quantum Electron.* **2005**, *11*, 149-156.

[31]. Y. Tian, K. L. Lee, and C. Lim, A. Nirmalathas, *J. Lightwave Technol.* **2017**, *35*, 4304-4310.

[32]. F. Tonolini and F. Adib, in Proc. Conf. ACM Special Interest Group Data Commun., *Networking across boundaries: enabling wireless communication through the water-air interface*, Budapest, Hungary, Aug. **2018**, pp. 117–131.

[33]. Y. C. Xin, Y. Li, V. Kovanis, A. L. Gray, L. Zhang, and L.F. Lester, *Opt. Express* **2007**, *15*, 7623-7633.

[34]. B. Shen, L. Chang, J. Liu, H. Wang, Q.-F. Yang, C. Xiang, R. N. Wang, J. He, T. Liu, and W. Xie, *Nature* **2020**, *582*, 365-369.

[35]. T. Erdogan, *J. Lightwave Technol.* **1997**, *15*, 1277-1294.

# *Supporting Information*

# Reciprocal Phase Transition Electro-optic Modulation

*Fang Zou,[1, †] Lei Zou,[2, †] Ye Tian,[3] Yiming Zhang, [4] Erwin Bente,[3] Weigang Hou,[5] Yu Liu,[4] Siming Chen,[2] Victoria Cao,[2] Lei Guo,[5] Songsui Li,[1] Lianshan Yan,[1] Wei Pan,[1] Dusan Milosevic,[3] Zizheng Cao,[3,6, *] A. M. J. Koonen,[3] Huiyun Liu,[2] and Xihua Zou[1, *]*

[1] School of Information Science and Technology, Southwest Jiaotong University, Chengdu 611756, China.
[2] Department of Electronic and Electrical Engineering, University College London, WC1E 6BT London, UK.
[3] Institute for Photonic Integration, Eindhoven University of Technology, Eindhoven 5600 MB, Netherlands.
[4] Institute of Semiconductors, Chinese Academy of Sciences, Beijing 100083, China.
[5] School of Communication and Information Engineering, Chongqing University of Posts and Telecommunications, Chongqing 400000, China.
[6] Peng Cheng Laboratory, Shenzhen 518055, China.

**Content:**

**Section S1. Qualitative differences between the travelling-wave model for a linear cavity and the DDE**

The carrier and field equations in the delay differential equation (DDE) model are the same as those in the travelling-wave model (TWM). However, these two models have different cavity geometries and propagations of the light field. The TWM has been used to numerically study a mode-locked laser with a linear cavity in which the light field is two counterpropagating waves [21, 22]. In contrast, a ring cavity in which the light field unidirectionally propagates is first assumed when the DDE model is used to analyze the dynamics of a mode-locked laser [19, 20, 23].

The differences between these two models mainly arise from deviations in the high-harmonic mode-locked regime due to the collision of multiple pulses. In the TWM, since the light field is two counterpropagating waves, the multiple pulses will collide with each other, resulting in faster saturation of the active section at the collision point. In addition, the interaction of a single pulse also occurs in the linear cavity, which is known as self-collision. The interaction between the forward and backward moving parts of a single pulse occurs at the end facets where the pulse is reflected. Similarly, the self-collision will lead to faster saturation if the collision point is located in the active section. In the DDE model, neither interaction nor self-collision occurs in the mode-locked laser with a ring cavity because the propagation of the light field is unidirectional. However, the qualitative dynamics, especially the bifurcation diagrams, for these two models are very similar when the mode-locked laser works in the anti-collision pulse-type mode [19].

In our simulation and experiment, the mode-locked laser works in the fundamental mode-locked regime in which only one pulse oscillates in the laser cavity within a round trip. In addition, since the saturable absorber (SA) and the gain elements in the mode-locked laser are placed away from the reflector at the end facets, as shown in **Figure 4**, self-collision in the active region does not occur. Therefore, the DDE model with a ring cavity is used to make qualitative predictions for experiments, in which the mode-locked laser has a linear cavity in the manuscript due to the higher simulation efficiency.

**Section S2. Theoretical analysis of phase transition EO modulation in the mode-locked laser based on the DDE model**

A monolithically integrated mode-locked laser is designed and implemented as a protype to reveal the phase transition EO modulation mechanism, shown in **Figure 2**. A theoretical analysis based on the DDE model is performed to investigate the dynamics of the mode-locked laser [19, 20, 23]. The laser cavity is considered to have a ring geometry, and propagation of the light field is unidirectional. Moreover, a distribution Bragg reflector (DBR) filter is introduced by the lumped approach, which serves as a reflector at the end facet.

The mode-locked laser with a ring cavity includes three sections and a DBR, as illustrated in **Figure S1**. Section I is a gain element, and Section III is a passive section. Section II is a reconfigurable element in which one side of the PN junction is connected to a switching port. When the switching port is set to be grounded, some of the oscillating optical energy transforms into an electrical voltage on Section II owing to the stimulated absorption such that Section II serves as an SA in the mode-locked laser cavity, as shown in **Figure S1a**. The mode-locked laser spontaneously evolves to the pulsed-wave (PW) state and emits pulse trains. In contrast, Section II serves as a passive section with linear loss when the switching port is triggered to be open-circuit, as shown in **Figure S1b**. The output of the mode-locked laser is continuous-wave (CW) light, which is determined as the CW state. When the electrical signal triggers transient actions through a control circuit to set the switching port as grounded or open-circuit, the phase transition modulation based on mode-locked laser is realized, as shown in **Figure S1c**.

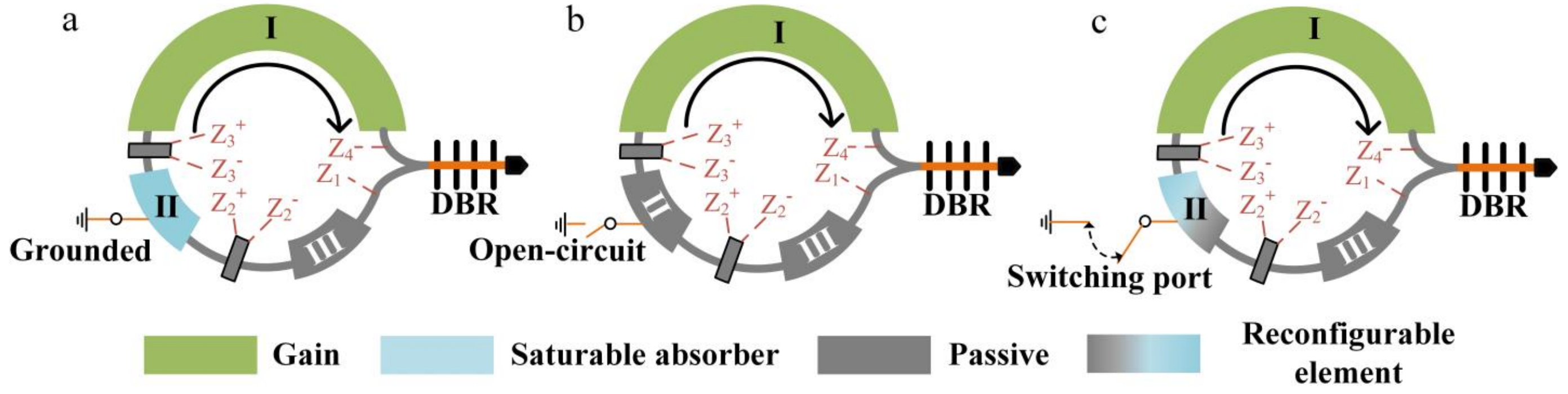

**Figure S1**. (a) PW state, (b) CW state and (c) phase transition modulation of the semiconductor laser with ring cavity.

### *S2.1. DDE model*

The field amplitude and carrier density equations can be derived from the standard TWM for semiconductor lasers:

$$\pm\frac{\partial\varepsilon_r^{\pm}(t,z)}{\partial z}+\frac{1}{v}\frac{\partial\varepsilon_r^{\pm}(t,z)}{\partial t}=\frac{g_r\Gamma_r}{2}(1-i\alpha_r)[n_r(t,z)-n_r^{tr}]\varepsilon_r^{\pm}(t,z), \tag{S1}$$

$$\frac{\partial n_r(t,z)}{\partial t}=\eta\frac{j_r(t,z)}{ed}-\gamma_r n_r(t,z)-vg_r\Gamma_r[n_r(t,z)-n_r^{tr}]\sum_{\pm}\left|\varepsilon_r^{\pm}(t,z)\right|^2, \tag{S2}$$

where r=g (or q or p) represents the gain element (or absorber or passive section), $v$ is the group velocity, $\Gamma_r$ is the optical confinement factor, $\alpha_r$ is the linewidth enhancement factor, and $g_r$ is the differential gain. The dynamical variables are the left (+) and right (−) propagating slowly varying complex electric field amplitudes $\varepsilon_r^{\pm}$ and the carrier density $n_r$. Because the ring laser cavity is under unidirectional operation in the DDE model, clockwise propagation is chosen, i.e., $\varepsilon_r^{+}=\varepsilon_r$ and $\varepsilon_r^{-}=0$. The field amplitude and carrier density of the gain, absorber, and passive sections are described in the comoving coordinates of $(t,z)\rightarrow(t',z')$ with $t'=t-z/v$ and $z'=z/v$.

For the passive section, since the carrier density $n_P(t,z)$ is constant and equal to zero, **Equation (S2)** is dismissed, and **Equation (S1)** can be written as:

$$\frac{\partial A_P(t',z')}{\partial z'}=0, \tag{S3}$$

where $A_p(t',z')=\sqrt{vg_g\Gamma_g}\varepsilon_p(t',z')$.

For the gain section, since currents are injected, the field amplitude and the carrier density can be written as:

$$\frac{\partial A_g(t',z')}{\partial z'}=\frac{1}{2}(1-i\alpha_g)N_gA_g(t',z'), \tag{S4}$$

$$\frac{\partial N_g(t',z')}{\partial t'}=\mathrm{J}_g(t',z')-\gamma_gN_g(t',z')-N_g(t',z')\left|A_g(t',z')\right|^2, \tag{S5}$$

where the field amplitude is defined as $A_g(t',z')=\sqrt{vg_g\Gamma_g}\varepsilon_g(t',z')$, $N_g(t',z')=vg_g\Gamma_g[n_g(t',z')-n_g^0]$ is the carrier density, and the pump parameter is defined as $\mathrm{J}_g(t',z')=vg_g\Gamma_g[\eta j_g(t',z')/(ed)-\gamma_gn_g^0]$.

There is no current injected into the SA section, i.e., $j_q(t,z)=0$, so its field amplitude and carrier density can be written as:

$$\frac{\partial A_q(t',z')}{\partial z'}=\frac{1}{2}(1-i\alpha_q)N_qA_q(t',z'), \tag{S6}$$

$$\frac{\partial N_q(t',z')}{\partial t'}=-\mathrm{J}_q-\gamma_qN_q(t',z')-r_sN_q\left|A_q(t',z')\right|^2, \tag{S7}$$

where the field amplitude and the carrier density of the SA are defined as $A_q(t',z')=\sqrt{vg_g\Gamma_g}\varepsilon_q(t',z')$ and $N_q(t',z')=vg_q\Gamma_q[n_q(t',z')-n_q^0]$. In addition, the pump parameter and the saturation energy ratio are defined as $\mathrm{J}_q=vg_q\Gamma_q\gamma_qn_q^0$ and $r_s=g_q\Gamma_q/g_g\Gamma_g$. The field amplitude and carrier density of each section are integrated by using the boundary conditions between them. Then, the differential equations that describe the PW state, CW state, and the phase transition modulation can be derived.

### *S2.2. PW state*

When the switching port is configured as grounded, Section II exhibits the feature of an SA, and the laser works in the PW state. The semiconductor laser consists of a gain element (i.e., Section I), an SA (i.e., Section II), a passive section (i.e., Section III), and a DBR filter, as shown in **Figure S1a**. First, the losses and feedback contributions are described above. Using the comoving coordinates and new definitions of $A_p(t',z')=\sqrt{vg_g\Gamma_g}\varepsilon_p(t',z')$, $A_g(t',z')=\sqrt{vg_g\Gamma_g}\varepsilon_g(t',z')$ ,and $A_q(t',z')=\sqrt{vg_g\Gamma_g}\varepsilon_q(t',z')$, the boundary conditions can be written as:

$$A_q(t,z_2^+)=\sqrt{k_1}A_p(t,z_2^-), \tag{S8}$$

$$A_g(t,z_3^+)=\sqrt{k_2}A_q(t,z_3^-), \tag{S9}$$

$$A_p(t'-T',z'_1+T')=\sqrt{k_3}f\left(t'-T'\right)*A_g(t'-T',z'_4), \tag{S10}$$

where $k_x = \alpha L_x$ is the linear loss of each section and $f(t)$ is the impulse response of the Bragg grating filter [35], which is introduced by the convolution "$*$" in **Equation (S11)**. For the comoving coordinates, the length of the round trip is expressed as $(t, z_1 + L) \to (t - (z_1 + L)/v,\ (z_1 + L)/v) \to (t' - T',\ z'_1 + T')$, in which $T'$ is the round-trip time in the cold cavity.

The field amplitude along each section can be calculated by integrating **Equations (S3), (S4)** and **(S6)**. The field amplitude can be written as:

$$A_P(t', Z_2^{'-}) = A_P(t', Z'_1), \quad \textbf{(S11)}$$

$$A_q(t', z_3^{'-}) = e^{-\frac{1}{2}(1-i\alpha_q)Q(t')} A_q(t', z_2^{'+}), \quad \textbf{(S12)}$$

$$A_g(t', Z_4^{'}) = e^{\frac{1}{2}(1-i\alpha_g)G(t')} A_g(t', Z_3^{'+}), \quad \textbf{(S13)}$$

where $G(t') = \int_{Z_2^{'+}}^{Z_3^{'-}} N_g(t', z') dz'$ and $Q(t') = -\int_{Z_1^{'}}^{Z_2^{'-}} N_q(t', z') dz'$ are defined as the dimensionless carrier densities integrated over the gain and absorber sections.

Considering the boundary conditions and field amplitude along each section, the field amplitude iterated over a round trip can be written as:

$$A_p(t' - T', z_1' + T') = f(t' - T') * R(t' - T') A_p(t' - T', z_1^{'}), \quad \textbf{(S14)}$$

where $R(t') = \sqrt{k} \exp\{0.5 \times [(1 - i\alpha_g)G(t') - (1 - i\alpha_q)Q(t')]\}$ is used to describe the net gain accumulated over one round trip. Here, we also define the total linear loss as $k = k_1 k_2 k_3$, and it represents all the non-resonant and mirror losses per round trip. By using the periodic boundary condition $A_p(t' - T', z'_1 + T') = A_p(t', z'_1)$, the evolution of the field over one round trip is expressed by the field amplitude in the passive section at position $z'_1$:

$$\frac{dA_p(t', z'_1)}{dt'} = \frac{df(t' - T')}{dt'} * R(t' - T') A_p(t' - T', z'_1). \quad \textbf{(S15)}$$

To model the differential equations for the time evolution of the carrier densities $G(t')$ and $Q(t')$, we integrate **Equations (S5)** and **(S7)** over the gain and absorber sections and obtain:

$$\frac{dG(t')}{dt'} = J_g(t', z') - \gamma_g G(t') - \int_{z_3^{'+}}^{z'_4} N_g(t', z') \left| A_g(t', z') \right|^2 dz', \quad \textbf{(S16)}$$

$$\frac{dQ(t')}{dt'} = J_q(t', z') - \gamma_q Q(t') - \int_{z_2^{'+}}^{z_3^{'-}} r_s N_q(t', z') \left| A_q(t', z') \right|^2 dz'. \quad \textbf{(S17)}$$

Here, $J_g(t', z') = \int_{z_2^{'+}}^{z_3^{'-}} \mathrm{J}_g(t', z') dz'$ and $J_q(t', z') = \int_{z_1^{'}}^{z_2^{'-}} \mathrm{J}_q(t', z') dz'$ are introduced, which represent the unsaturated gain and absorption parameters.

**Equations (S12)** and **(S13)** are multiplied by the complex conjugate of the field amplitudes $A_g^*(t',z')$ and $A_q^*(t',z')$ to calculate the coupling between the field amplitude and carrier density. Then, by adding the complex conjugate of the equation by multiplied by $A_g(t',z')$ and $A_q(t',z')$, we can obtain:

$$\left|A_q(t',z'^{-}_3)\right|^2-\left|A_q(t',z'^{+}_2)\right|^2=\int_{z'^{+}_2}^{z'^{-}_3}N_q\left|A_q(t',z')\right|^2dz', \tag{S18}$$

$$\left|A_g(t',z'_4)\right|^2-\left|A_g(t',z'^{+}_3)\right|^2=\int_{z'^{+}_3}^{z'_4}N_g\left|A_g(t',z')\right|^2dz'. \tag{S19}$$

The terms on left side can be expressed in terms of $A(t')$, $G(t')$, and $Q(t')$ by introducing the boundary conditions and integrating the response of each section:

$$k_1 r_s(e^{-Q(t')}-1)\left|A_P(t',z_1^{'})\right|^2=\int_{z'_1}^{z'^{-}_2}r_s N_q\left|A_q(t',z')\right|^2dz', \tag{S20}$$

$$k_1k_2e^{-Q(t')}(e^{G(t')}-1)\left|A_p(t',z_1^{'})\right|^2=\int_{z'^{+}_2}^{z'^{-}_3}N_g\left|A_g(t',z')\right|^2dz'. \tag{S21}$$

After rescaling the field amplitude and the saturation energy ratio by $A(t')=\frac{1}{\sqrt{k_1k_2}}A_p(t',z_1^{'})$ and $\hat{r}_s=\frac{r_s}{k_2}$, the DDE model for the semiconductor laser with a grounded switching port is summarized as:

$$\frac{dA(t')}{dt'}=\frac{df\left(t'-T'\right)}{dt'}*R\left(t'-T'\right)A(t'-T'), \tag{S22}$$

$$\frac{dG(t')}{dt'}=J_g-\gamma_g G(t')-e^{-Q(t')}(e^{G(t')}-1)\left|A(t')\right|^2, \tag{S23}$$

$$\frac{dQ(t')}{dt'}=J_q-\gamma_q Q(t')-\hat{r}_s(e^{-Q(t')}-1)\left|A(t')\right|^2, \tag{S24}$$

$$R(t')=\sqrt{k}e^{\frac{1}{2}[(1-i\alpha_g)G(t')-(1-i\alpha_q)Q(t')]}. \tag{S25}$$

## *S2.3. CW state*

When the switching port is open-circuit, Section II is a passive element, and the laser works in the CW state. In this context, the semiconductor laser consists of a gain section (i.e., Section I), two passive sections (i.e., Sections II and III), and a DBR filter, as shown in **Figure S1b**. For the CW-state model, the boundary condition, field, and carrier equations are the same as those in the PW state except for those of Section II. The field amplitude and carrier density of Section II in the CW-state model can be expressed as:

$$A_p(t',z'^{-}_3)=A_p(t',z'^{+}_2), \tag{S26}$$

$$\int_{Z'^{+}_2}^{Z'^{-}_3}N_P(t',z')dz'=0. \tag{S27}$$

Then, following the derivation for the PW state, the DDE model for the semiconductor laser with an open-circuit switching port can be summarized as:

$$\frac{dA(t')}{dt'} = \frac{df\left(t'-T'\right)}{dt'} * R\left(t'-T'\right) A(t'-T'), \quad \text{(S28)}$$

$$\frac{dG(t')}{dt'} = J_g - \gamma_g G(t') - (e^{G(t')} - 1)\left|A(t')\right|^2, \quad \text{(S29)}$$

$$R(t') = \sqrt{k} e^{\frac{1}{2}(1-i\alpha_g)G(t')}. \quad \text{(S30)}$$

### *S2.4. Phase transition EO modulation based on a mode-locked laser*

The phase transition EO modulation is implemented with and demonstrated by a monolithically integrated multi-section mode-locked laser protype. Here, a new method to adjust the control parameter's value is discovered, namely the alternative transparency and saturable absorption effects in a reconfigurable element (i.e., Section II) as shown in **Figure S1c**. The control parameter manipulated by the reconfigurable element is involved in the absorption parameters and the carrier recovery rate corresponding to $J_q(t)$ and $\gamma_q(t)$ in the DDE model. Following the derivation, the dynamical model of the phase transition EO modulation based on mode-locked laser is described in **Equations (S31)-(S34)**.

$$\frac{dA(t')}{dt'} = \frac{df\left(t'-T'\right)}{dt'} * R\left(t'-T'\right) A(t'-T'), \quad \text{(S31)}$$

$$\frac{dG(t')}{dt'} = J_g - \gamma_g G(t') - e^{-Q(t')}(e^{G(t')} - 1)\left|A(t')\right|^2, \quad \text{(S32)}$$

$$\frac{dQ(t')}{dt'} = J_q(t') - \gamma_q(t')Q(t') - \hat{r}_s(e^{-Q(t')} - 1)\left|A(t')\right|^2 + L_Q \frac{dq_0(t')}{dt'}, \quad \text{(S33)}$$

$$R(t') = \sqrt{k} e^{\frac{1}{2}[(1-i\alpha_g)G(t') - (1-i\alpha_q)Q(t')]}. \quad \text{(S34)}$$

Here, **Equation (S33)** has been expanded with the term $L_Q[{}^{dq_0(t')}\!/_{dt'}]$, which originates from the time dependency of the unsaturated absorption in the linear approximation between the carrier density and the absorption.

For the InP quantum-well SOA, the pump parameter $J_g$ linearly increases with the injected current $I_{SOA}$, as depicted in **Figure S2a**. The relationship between the current and pump parameters can be derived as:

$$J_g(t', z') = L_g g_g \Gamma_g \left[\frac{I_{SOA}}{e\sigma_{xy} L_g} - \gamma_g n_g^{tr}\right], \quad \text{(S35)}$$

where $L_g$, $g_g$, and $\Gamma_g$ are the gain element length, the differential gain, and the confinement coefficient for the SOA gain. The carrier injection rate is determined from $I_{SOA}/(e\sigma_{xy}L_g)$ with injected current $I_{SOA}$, elementary charge $e$, and active area $\sigma_{xy}$. $n_g^{tr}$ is the gain element carrier transparency density.

The phase transition modulation based on the mode-locked laser is realized by trigging the switching port to be grounded or open-circuit via a control circuit. On the one hand, when the switching port is open-circuit, Section II is configured as a passive element, so the unsaturated absorption $J_q(t)$ constantly equals zero. Since the right part of **Equation (S33)** is negative after the transition time, the carrier density in Section II, $Q(t')$, decreases to null. After spontaneously evolving to the self-sustaining CW state, the DDE model is simplified to the CW state described by **Equations (S28)-(S30)** and includes only the two coupled equations of $A(t')$ and $G(t')$. On the other hand, Section II behaves as an SA if the switching port is set as grounded. $J_q(t) = L_Q q_0 / t_Q$ and $\gamma_q(t) = 1/t_Q$ can be calculated by setting the bias voltage to zero. Based on the experimental values, $t_Q$ and $q_0$ of the SA are fitted as empirical expressions:

$$t_Q(t) = 85 \times e^{-\frac{V_{bias}(t)-1.6}{2.1}}, \quad \text{(S36)}$$

$$q_0(t) = 95.71 \times \tanh[0.84(V_{bias}(t) - 2.44)] + 182.12, \quad \text{(S37)}$$

where $t_Q$ is expressed as an exponential function with the revised bias voltage [24], and $q_0$ can be approximated by a hyperbolic tangent [25], shown in **Figures S2a** and **b**.

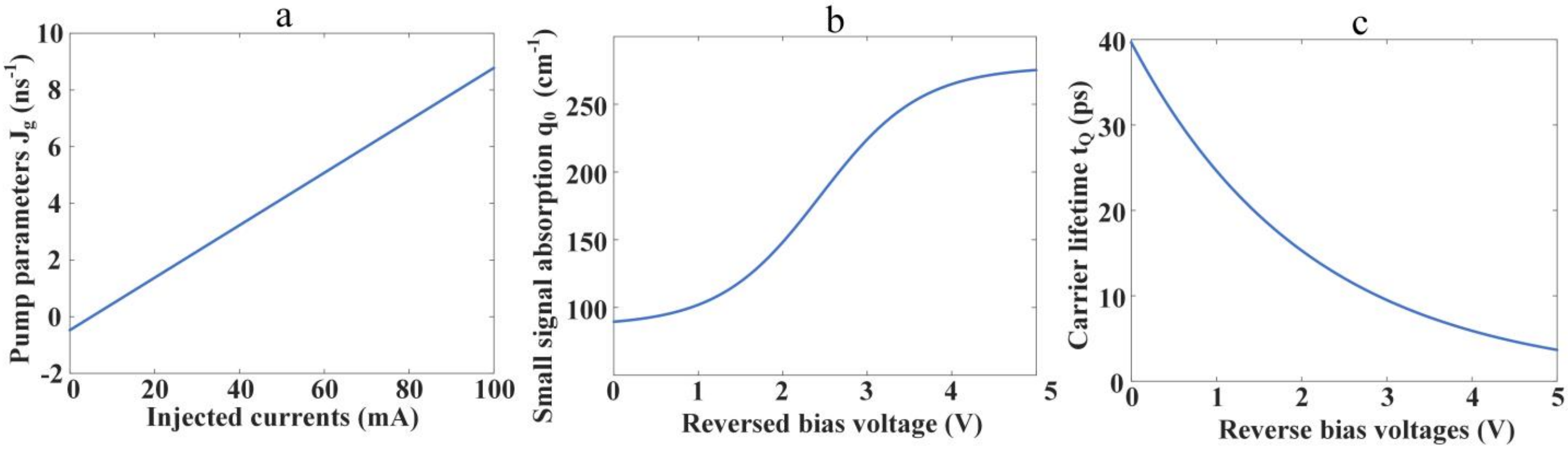

**Figure S2**. (a) Pump parameter $J_g$, (b) small signal absorption $q_0$, and (c) carrier lifetime $t_Q$ obtained using the DDE model.

## Section S3. Parameters, processes, and results of the simulation

In the simulation, the fourth-order Runge–Kutta algorithm with a time resolution of $\Delta t = T_r / 2^8 \approx 140.6\ fs$ is used to solve the differential equations. Since the spontaneous recombination of carriers in the SOA induces random noise in both the amplitude and phase, a complex random process is used in the simulation. The semiconductor laser is self-starting from the noise, and then, it reaches the steady state after some round trips. Some empirical expressions and experimental values for InP materials extracted from references [24-29] are used in the simulation, as listed in **Table S1**.

A 200-μm Bragg grating in length is implemented as the DBR in multi-section mode-locked laser. According to the parameters in **Table S1**, the reflected spectrum $f(\omega)$ is shown in **Figure S3a**. The reflection of the DBR (i.e., Bragg grating) is approximately 0.58, and the FWHM is 262 GHz, which can support at least the 10th harmonic frequency in the PW state. After inverse Fourier transforms, we can obtain the impulse response of the Bragg grating $f(t)$, and the impulse response is shown in **Figure S3b**. The amplitude of the DBR attenuates to zero after approximately 10 ps. Thus, we assume that the filter has a periodic response, in which the period is equal to the round-trip time of the laser (i.e., 36 ps) in the simulation. $A(t')$ is the field amplitude of the oscillation in the laser cavity. The transmission of the DBR is the real output of the semiconductor laser, which equals the convolution between $A(t')$ and the transmission response of the DBR.

**Table S1.** Parameters for the semiconductor mode-locked laser

| Parameter | Symbol | Value | Ref. |
|---|---|---|---|
| InP semiconductor laser | | | |
| Round-trip time in the cold cavity | $T_r$ | 36 ps | - |
| Length of Section II (absorber or passive) | $L_{\mathrm{II}}$ | 60 μm | - |
| Length of Section I (gain) | $L_{\mathrm{I}}$ | 750 μm | - |
| Non-resonant and mirror losses | $k$ | 0.116 | - |
| Linewidth enhancement factor of gain element | $\alpha_g$ | 3 | [27] |
| Linewidth enhancement factor of absorber | $\alpha_q$ | 3 | [27] |
| Saturation energy ratio | $\hat{r}_s$ | 25 | [28] |
| Carrier recovery rate of the gain element | $\Upsilon_g$ | 0.63 $ns^{-1}$ | [28] |
| Injected current | $I_{SOA}$ | 60 mA | - |
| Confinement factor | $\Gamma$ | 0.074 | [29] |
| Active area | $\sigma_{xy}$ | 0.0624 $\mu m^2$ | [29] |
| Differential gain | $a_{NG}$ | $1.25\times10^{-20}$ $m^2$ | [26] |
| Elementary charge | $e$ | $1.6\times10^{-19}$ C | - |
| Amplified spontaneous emission (ASE) noise spectral density | $N_0$ | $10^{-20}$ | [26] |
| Carrier lifetime | $t_Q$ | **Equation (S36)** | [24] |
| Small signal absorption | $q_0(t)$ | **Equation (S37)** | [25] |
| Bragg grating | | | |
| Grating length | $L_{Grating}$ | 200 μm | [35] |
| Coupling coefficient | $\kappa$ | 50 $cm^{-1}$ | [35] |
| Effective index | $n_{eff}$ | 3.38 | [35] |

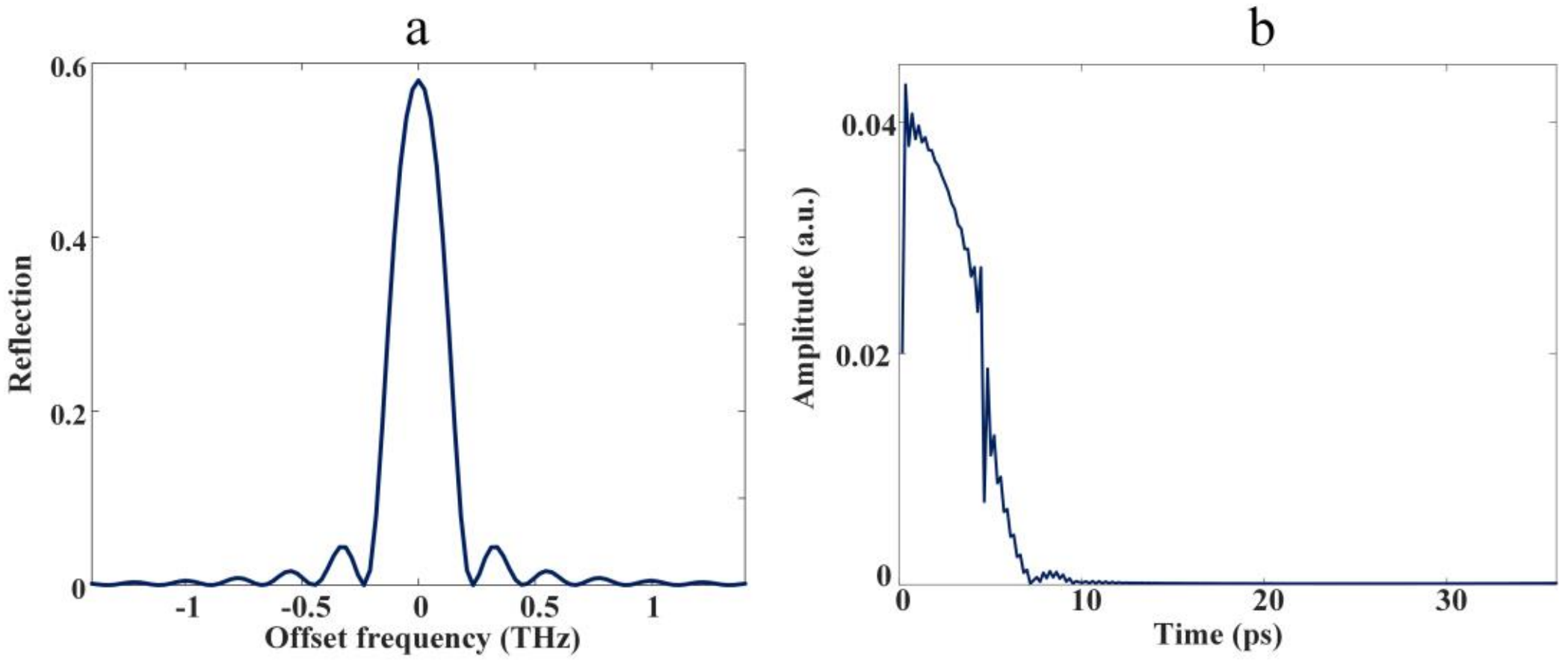

**Figure S3**. (a) Reflection spectrum and (b) corresponding impulse response of the DBR (i.e., Bragg grating).

When the switching port is configured as grounded, $J_q$ and $\gamma_q$ can be calculated by setting the reverse bias as 0 V. By using the DDE model expressed in **Equations (S22)-(S25)**, the field amplitude, electrical spectrum and optical spectrum of the PW state are displayed in **Figure S4**. After a relaxation time, the semiconductor laser reaches a self-sustaining PW state. Pulse light trains with a period of 59 ps are generated, as shown in **Figure S4a** and the magnified view in **Figure S1b**. The corresponding optical and electrical spectra are frequency combs with a frequency spacing of 18 GHz, as shown in **Figures S4c** and **d**.

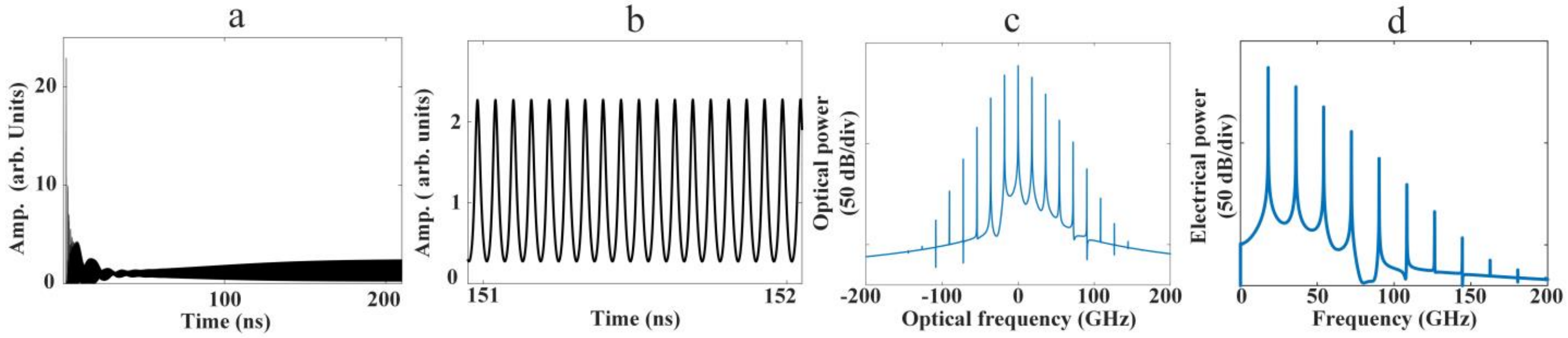

**Figure S4.** Simulation results of the PW state. (a) Field amplitude in the time domain. (b) Zoom-in view of the field amplitude from 151 ns to 152 ns. (c) Optical and (d) electrical spectra of the PW state.

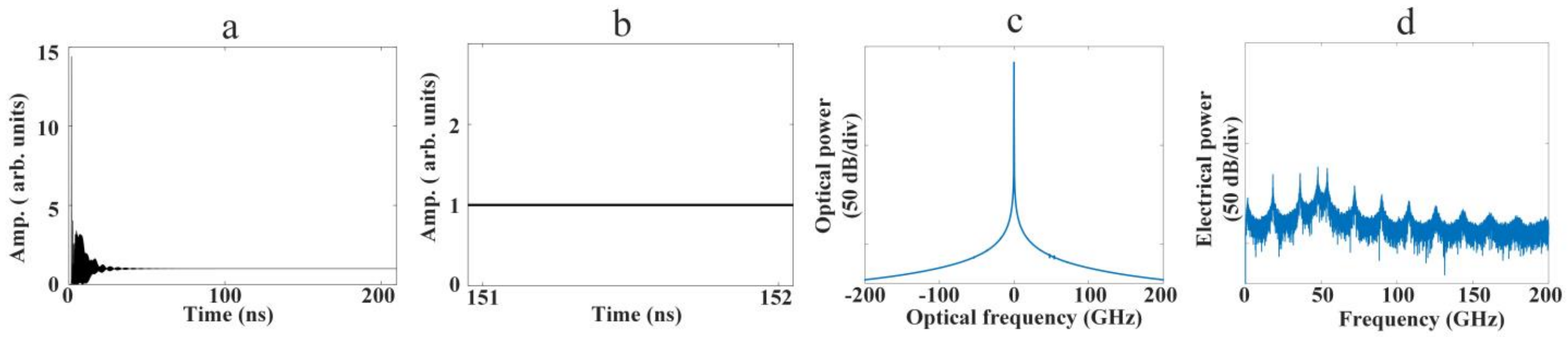

**Figure S5**. Simulation results of the CW state. (a) Field amplitude in the time domain. (b) Zoom-in view of amplitude from 151 ns to 152 ns. (c) Optical and (d) electrical spectra of the CW state.

The mode-locked laser works in the CW state, and the model can be described by **Equations (S28)-(S30)**, if the switching port is open-circuit. The field amplitude, spectrum, and optical spectrum of the CW state are shown in

**Figure S5**. A continuous wave is generated after a relaxation time, as shown in **Figures S5a** and **b**. In the optical spectrum, a single mode is selected so that a CW light is emitted. After photodetection and DC blocking, the electrical spectrum shows noise and side modes with low power, as shown in **Figures S5c** and **d**.

When the driving electrical signal triggers transient actions via a control circuit to set the switching port between open-circuit (‘0’) and grounded (‘1’), the mode-locked laser spontaneously evolves to the CW or PW state. Therefore, the phase transition EO modulation is accomplished by the mode-locked laser, as depicted in **Figure S6a**. Since the parameters $J_q$ and $\gamma_q$ vary with the driving electrical signal as aforementioned, the dynamical model of the phase transition EO modulation described in **Equations (S31)-(S34)** is simplified to the model of the CW or PW state. The simulation results of phase transition modulation perfectly match the CW or PW state according to the driving electrical signal, as shown in **Figures S6b** and **c**. In addition, the period of the coded signal is 294.912 ns and the transition times are approximately 200 ns and 40 ns for the spontaneous CW-PW and PW-CW transitions, respectively.

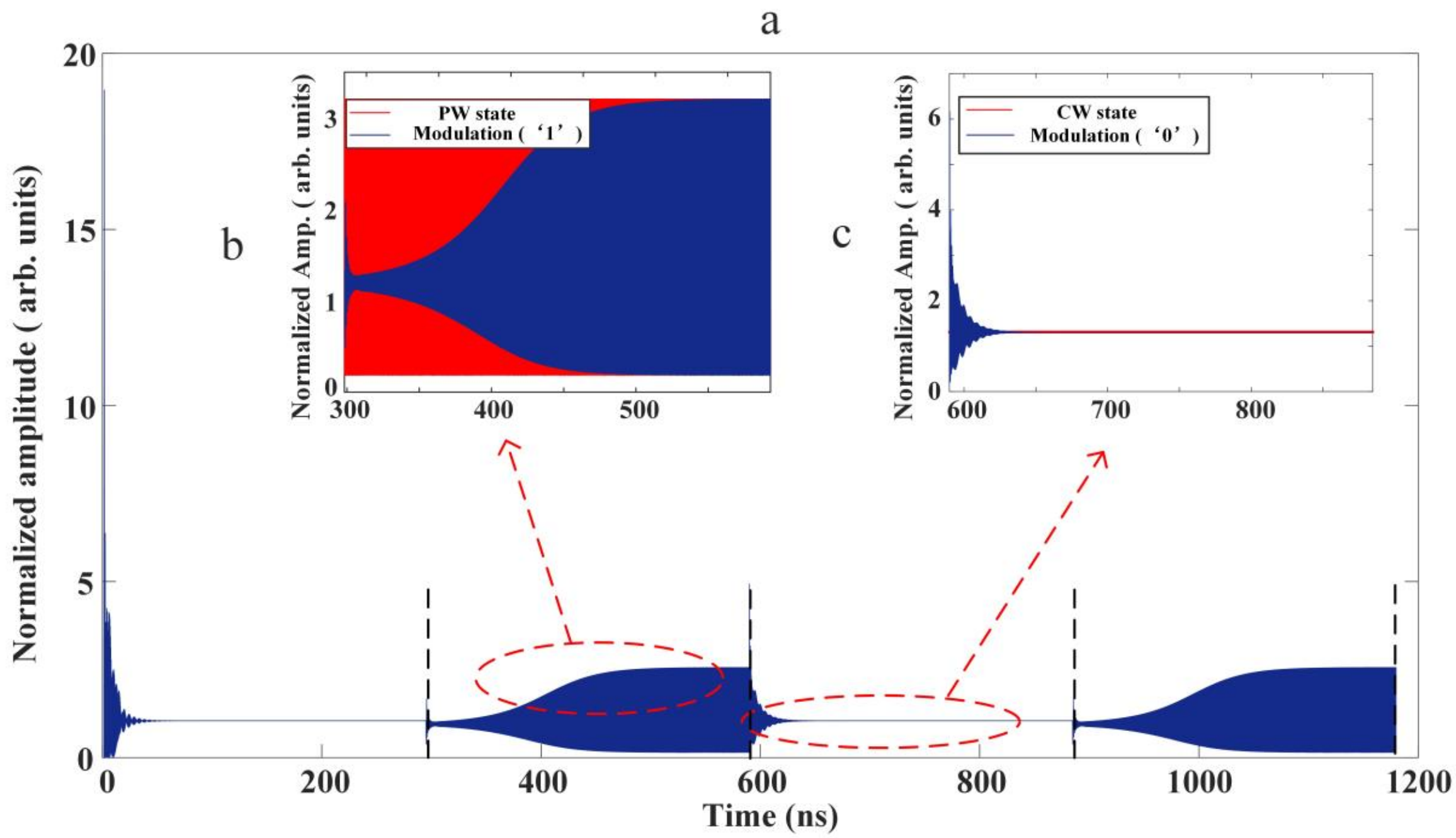

**Figure S6**. (a) Qualitative dynamics of spontaneous CW-PW and PW-CW state transitions. Zoom-in view of (a) the CW-PW state transition and (c) the PW-CW state transition

## Section S4. Calculation of modulation energy efficiency

A MOSFET switch (ALD110900) with a threshold voltage of 20 mV and an input capacitance of 2.5 pF serves as the control circuit to convert the electrical driving signal into the grounded or open-circuit action. The modulation energy efficiency is defined as the energy consumption per bit during the phase transition modulation, and it can be estimated by the driving power dissipation of the gate in phase transition modulation. The PN junction of the MOSFET has no leakage current (<200 pA) essentially, and the transitions between '0' and '1' are equally distributed in an NRZ PRBS. Thus, the modulation energy efficiency can be simplified as [11]:

$$\text{Energy/bit} = C_{ISS} \times (V_G)^2 / 4, \tag{S38}$$

where $C_{ISS}$ is the input capacitance of the MOSFET, and $V_G$ is the voltage on the gate. Therefore, the modulation energy efficiency is calculated to be 3.06 fJ/bit, according to the driving electrical signal with a peak-to-peak voltage of 70 mV. The capacitance of the adopted MOSFET can be reduced further to gain a higher modulation efficiency in future.